# FPScreen: A Rapid Similarity Search Tool for Massive Molecular Library Based on Molecular Fingerprint Comparison

[1]Lijun Wang, [1]Jianbing Gong, [1]Yingxia Zhang, [2]TianMou Liu, [1,*]Junhui Gao
[1]Zhangjiang Center for Translational Medicine, Biotecan Medical Diagnostics Co., Ltd. Shanghai, China
[2]University at Buffalo, Mathematics Department, Buffalo, New York, USA.
*Email: jhgao@biotecan.com, jhgao68@163.com

**Abstract**

We designed a fast similarity search engine for large molecular libraries: FPScreen. We downloaded 100 million molecules' structure files in PubChem with SDF extension, then applied a computational chemistry tool RDKit to convert each structure file into one line of text in MACCS format and stored them in a text file as our molecule library. The similarity search engine compares the similarity while traversing the 166-bit strings in the library file line by line. FPScreen can complete similarity search through 100 million entries in our molecule library within one hour. That is very fast as a biology computation tool. Additionally, we divided our library into several strides for parallel processing. FPScreen was developed in WEB mode.

## 1 Introduction

Molecular similarity search is a virtual screening method for targeting objective molecules in molecule libraries. It is widely used in drug screening [1] and compound recognition [2]. It saves a great deal of labor and cost.

There are a number of molecular similarity searching methods, such as molecular fingerprint search, 3D structure search [3] and molecular electrostatic potential arrangement search [4]. Molecular fingerprint search proceeds as follows: recognizing specific sub-structures in the molecular structure, transforming the molecular structure into series of binary fingerprints sequences, comparing the similarities of fingerprint sequences to find the molecules with significant similarities. When we are proceeding with 3D structure search algorithm, we use three-dimensional molecular descriptors to calculate various indicators of molecular similarity, and measure the significances. For electrostatic potential alignment search algorithm, we first aligns the target molecules, superimpose their similar structures, and calculate the electrostatic potential. Then it is transformed into a similarity index of -1 to +1. Finally, we can acquire the molecules with great similarities.

At present, molecular fingerprint search is the most simple and widely used tool for molecular similarity search for its efficiency and specialty.

We presents a fast search engine for molecular fingerprint database, FPScreen.

## 2 Results

### 2.1 User Interface

The molecular fingerprint search engine's user interface page is shown in Figure 1. The first half of the title is "Molecular Fingerprint Big Data Search Engine". Under the title is the input box for molecular fingerprint search content. Enter the search content and click the search button to see the search proceed as the progress bar in Figure 1. Under the search text box is the number of molecules in the compound totaling 95417114. We divided the dataset into three strides, dataset A, dataset B and dataset C for parallel processing. Different datasets are displayed in different color progress bars.

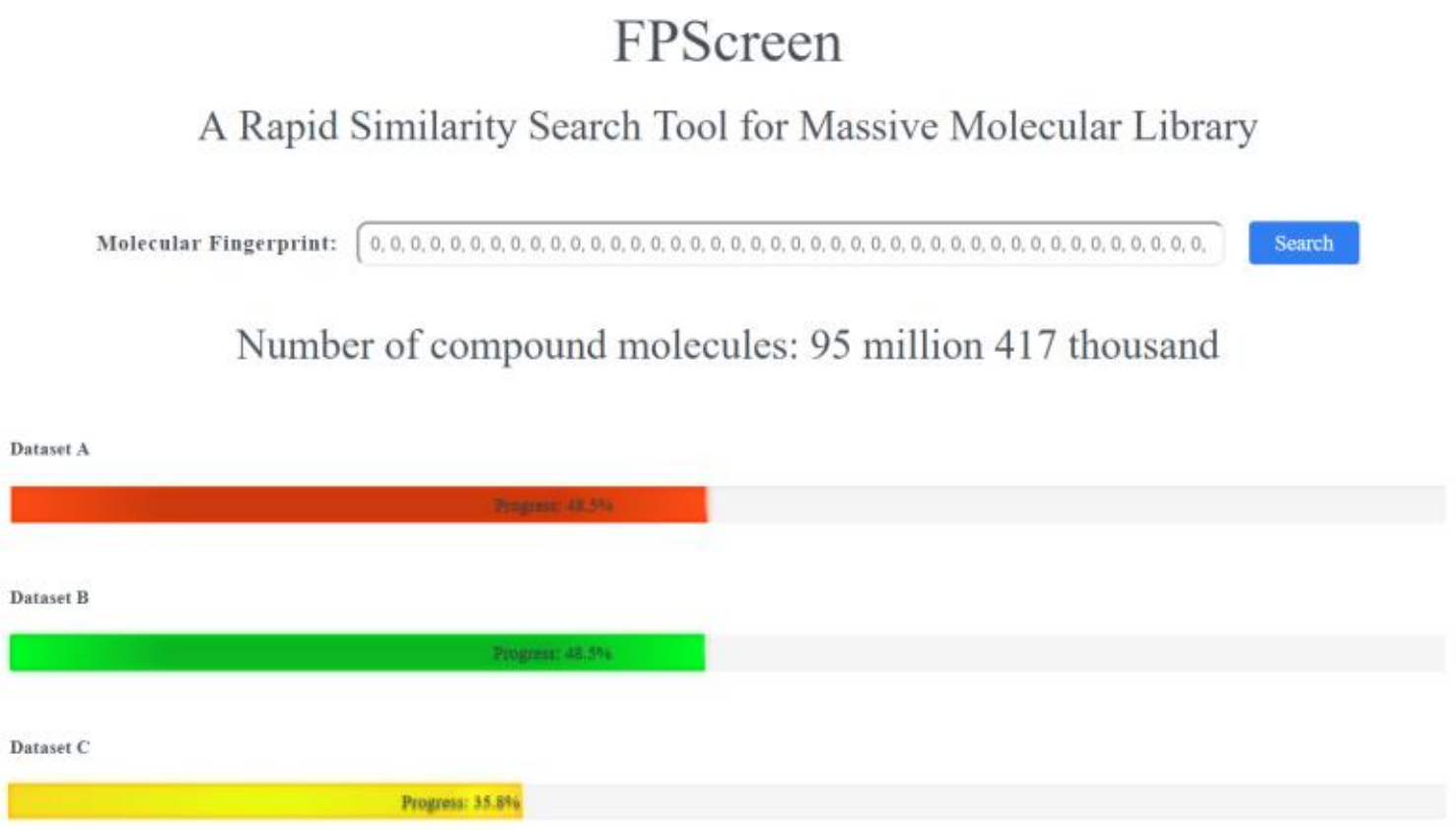

Figure 1　User Interface for Similarity Search

The search results is shown in Figure 2 displaying the top 30 search results. The results include the Manhattan distance between PubMed CID and the target molecule . The results are arranged in ascending order of distance.

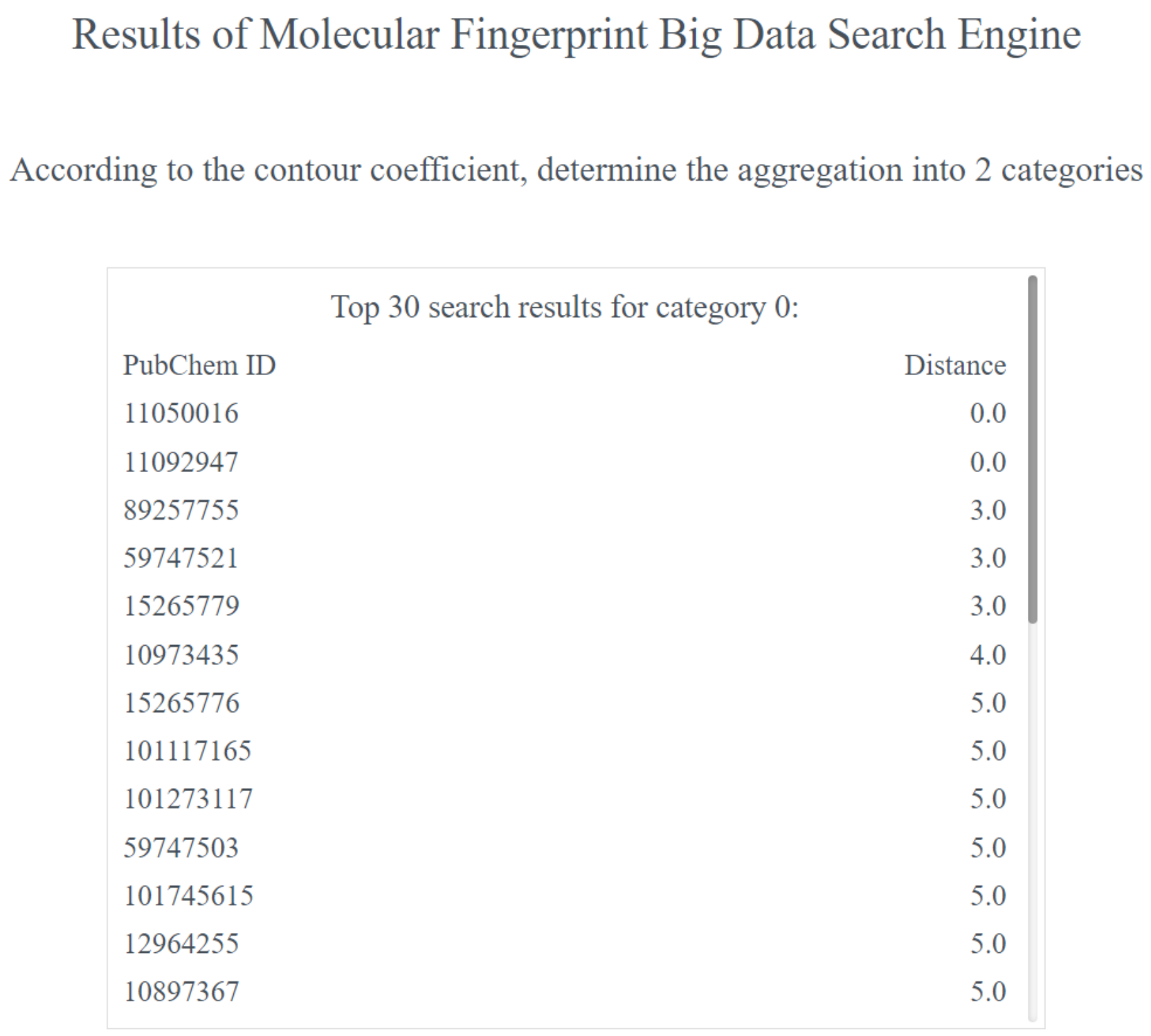
Results of Molecular Fingerprint Big Data Search Engine

According to the contour coefficient, determine the aggregation into 2 categories

Top 30 search results for category 0:

| PubChem ID | Distance |
|---|---|
| 11050016 | 0.0 |
| 11092947 | 0.0 |
| 89257755 | 3.0 |
| 59747521 | 3.0 |
| 15265779 | 3.0 |
| 10973435 | 4.0 |
| 15265776 | 5.0 |
| 101117165 | 5.0 |
| 101273117 | 5.0 |
| 59747503 | 5.0 |
| 101745615 | 5.0 |
| 12964255 | 5.0 |
| 10897367 | 5.0 |

Figure 2　Pages showing search results

**2.2 Search Speed**

There are 95417114 molecules in the library. It takes about 40 minutes to complete a similarity search. Our test computer has the following configuration: CPU: i5-6500, 3.20GHz; RAM: 16.0GB.

If we use RDKit package to complete the same molecular similarity search task, it will take more than 5 days. We increased the search speed by two orders of magnitude using our search engine.

**3 Discussion**

### 3.1 Integration Mode

Due to the huge amount of data processed by the molecular fingerprint search system, it naturally poses a huge challenge for the computer computing requirements. We consider two integration schemes in accordance to the actual requirements of the software system, hardware and practical scenarios. The first is to deploy user operation part and computing part on Windows and Linux servers respectively, and interact with each other by calling interfaces. The advantage is that it is technically easy to implement and loosely coupled with each other, but the disadvantage is that it cannot achieve real-time interaction and return to the intermediate computing process. The second scheme is that the two systems interact with each other in a queuing way. The advantage is that the two systems can decouple from each other and have no dependence on each other. It does not need to wait for the other side to complete before proceeding to the next step. In addition, the asynchronous operation between systems is achieved, one does not interfere with each other. This scheme is also not affected by the long calculation time. Although this scheme has many advantages, it is more complex to implement, which will inevitably increase the cost of development. Considering comprehensively, we still choose the second scheme, which highlights the advantages of our system, and is more compatible to future expansion and stability.

### 3.2 Multithread and Single Thread

At present, the background algorithm uses single-threaded computing. The main advantage of single-threaded processing lies in the sequential processing of tasks. For multi-threaded applications, the system is stable, extensible, rich in software options, and the development of synchronous applications is relatively easy. However, the efficiency of single-threaded processing is usually lower than that of multi-threaded applications because it needs to start new tasks only after the last task is completed. The processing time is longer. Multithreading can run multiple processes at the same time, which can significantly improve the performance of multitasking. Therefore, we will use multithreading in future upgrades to improve the computing speed.

### 3.3 Accuracy of Search Results

We randomly selected a molecule, PubChem CID: 11050016. After our tool search, PubChem CIDs of the closest molecule were 11092947 (distance: 0.0), 89257755 (distance 3.0), 59747521 (distance 3.0), 15265779 (distance 3.0). The information of the target molecule and the two closest molecule is shown in the table 1 below.

Table 1 Information about the closest two molecules

| **PubChem CID:** | 11050016 | 11092947 | 89257755 |
|---|---|---|---|
| **Structure:** | 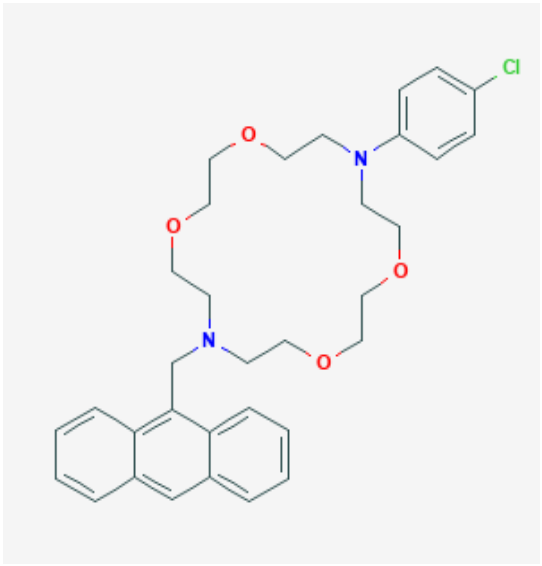 | 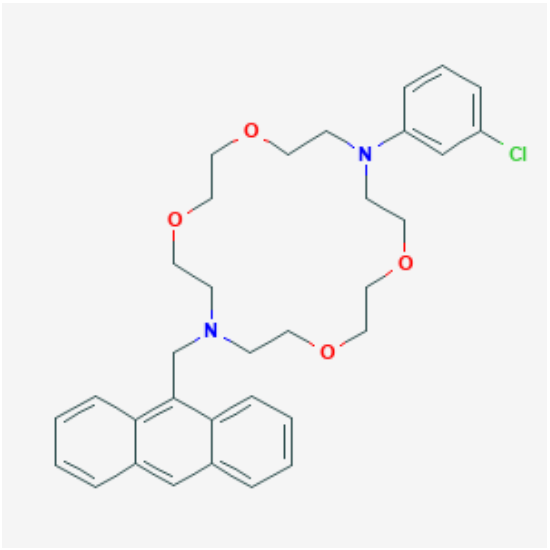 | 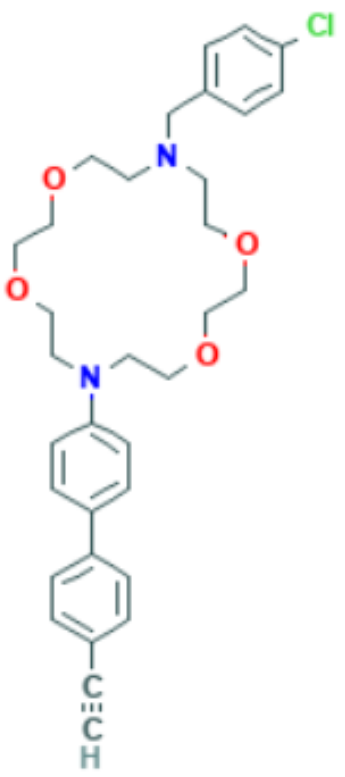 |
| **InChI Key:** | GBDBPEMLTAESAO-UHFFFAOYSA-N | PGMNXQKPIHGADU-UHFFFAOYSA-N | XKHSYBVSKSHOKT-UHFFFAOYSA-N |

| Molecular Formula: | $C_{33}H_{39}ClN_2O_4$ | $C_{33}H_{39}ClN_2O_4$ | $C_{33}H_{39}ClN_2O_4$ |
|---|---|---|---|
| Chemical Names: | 1-(4-Chlorophenyl)-10-(9-anthrylmethyl)-1,10-diaza-4,7,13,16-tetraoxacyclooctadecane | 1-(3-Chlorophenyl)-10-(9-anthrylmethyl)-1,10-diaza-4,7,13,16-tetraoxacyclooctadecane | SCHEMBL14140523 |
| Molecular Weight: | 563.135 g/mol | 563.135 g/mol | 563.135 g/mol |

The former CID 11050016 connected the large ring with a single benzene ring, and the fourth C connected the N atom counterclockwise. The third C connected the N atom at CID 11092947. The Manhattan distance between the two molecules was 0, indicating that the two molecules were very close. At the same time, the 166-dimensional MACCS fingerprint could not completely distinguish the differences between the molecules. But we can see from the results that our results are more accurate.

## 4 Materials and Methods

### 4.1 Data

We obtained small molecule compounds structure data from PubChem [5]. PubChem is a database of chemical modules, which is supported by the National Institutes of Health and maintained by the National Biotechnology Information Center.

We downloaded the structure files of all compounds by June 2018 through FTP, totaling about 100 million.

### 4.2 Molecular Fingerprint Extraction

There are many kinds of molecular fingerprints. We chose MACCS Key [5]. MACCS Key is the earliest and most popular molecular fingerprint developed by the former MDL [7] [8]. The public version contains 166 bonds (166 0 or 1), each of which corresponds to a specific molecular feature. Table 2 lists the specific information of the 1st to 24th molecular fingerprints [9].

Table 2 The specific meanings of the 1st to 24th molecular fingerprints

| | | |
|---|---|---|
| 1 ISOTOPE | 9 GROUP VIII (Fe...) | 17 CTC |
| 2 103 < ATOMIC NO. < 256 | 10 GROUP IIA (ALKALINE EARTH) | 18 GROUP IIIA (B...) |
| 3 GROUP IVA,VA,VIA PERIODS 4-6 (Ge...) | 11 4M RING | 19 7M RING |
| 4 ACTINIDE | 12 GROUP IB,IIB (Cu...) | 20 SI |
| 5 GROUP IIIB,IVB (Sc...) | 13 ON(C)C | 21 C=C(Q)Q |
| 6 LANTHANIDE | 14 S-S | 22 3M RING |
| 7 GROUP VB,VIB,VIIB (V...) | 15 OC(O)O | 23 NC(O)O |
| 8 QAAA@1 | 16 QAA@1 | 24 N-O |

RDKit [10] is an open source chemical informatics and machine-learning toolkit. We use RDKit to extract MACCS166 molecular fingerprints of compound structure files.

### 4.3 Similarity Search

By calculating the Manhattan distance between the two molecular fingerprints, we can show the similarity between the two molecules. The formula for Manhattan distance is as follows:

$$dist_{man}(x,y) = \sum_{j=1}^{n} \left| a_j - b_j \right|$$

In the above formulas, A and B are two molecules, j is the indicator for characteristic and n equals 166.

At the same time, in order to improve the operation speed, we adopted the subtraction operation of matrices and vectors. Subtracting a matrix from a vector produces another matrix, $C_{i,j} = A_{i,j} - b_j$ (where i is the $i^{th}$ row of the matrix and j is the $j^{th}$ column of the matrix). That is subtracting each row of A by vector b. This is called broadcasting. Here, matrix A is a matrix composed of molecular fingerprints (166 columns) of multiple molecules (rows), and vector B is a molecular fingerprint (166 features) of a single molecule. C is the resultant matrix. By summing the absolute values of each row of the result matrix, we have the Manhattan distance between each molecule in the library and the target molecule.

Input: Target molecule MACCS166, molecule library file F containing MACCS166
Output: PubChem CID with the smallest difference between the molecular fingerprints of N and the target molecule, and distance

Technological process:
1) Obtain all the molecules from F, construct CID, matrix A
2) Target molecule. Molecular fingerprint generates vector B.
3) $C_{i,j} = A_{i,j} - b_j$
4) K = SUM (ABS (C)), Manhattan distance, summing absolute value of each row in C to get K
5) CID + K
6) Sort, take the first N in the ascending order

Figure 3 Similarity Search Algorithm

If we are parallelizing multiple files, the results of all searches will be combined and sorted.

**4.4 Development Ideas of Application Website**

Molecular fingerprint search system is divided into two parts: user operation and calculation.

User operation part adopts B/S architecture mode, users can access the system directly through the browser, eliminating the tedious installation. The system is deployed on Windows server; IIS serves as Web server, and the system is developed in ASP.NET MVC [11] framework. We use the most popular RESTful architecture [12] mode as our back-end. It is a lightweight, cross-platform, cross-language architecture design, which facilitates the communication between different front-end devices and back-end. It can provide services for Web, iOS and Android through a unified interface. The front-end uses Html5 [13], CSS3 [14], Bootstrap [15], Jquery and other languages and frameworks, which are concise and flexible, making Web development faster.

User interface mainly focuses on two functions of "single molecular search" and "batch clustering search". "Single molecular search" refers to the search for a single molecular fingerprint, using Manhattan distance algorithm to find the top 30 molecules; while "batch clustering search" refers to a batch of molecular fingerprints, using clustering algorithm to search through the fingerprint database and find the top 30 molecules.

About the computation, considering that the molecular fingerprint search system needs to process huge amounts of data, the hardware requirements such as processing power and memory are quite high, so we deployed the computing part in the Linux server environment, making use of its advantages of multi-threading, high concurrency and stability. We used Python [16] [17] language for computation and to carry out artificial intelligence algorithm, considering its rich and powerful open source library and data analysis advantages.

Because there could be different system platforms, the user operation part and the computing part of the system may not directly interact with each other. We used queuing scheme as the bridge to enable real-time and asynchronous interaction between systems.

## 5 Concluding remarks

The purpose of virtual screening is to screen new approximate compounds from dozens or even hundreds of millions of molecules. This method does not consume samples, reduces the cost of screening, does not have sample restrictions, shortens the research cycle and reduces the research and development costs.

When the molecule pool's size reaches tens of millions or even billions, we need a fast engine. FPScreen is developed for this purpose. Tests show that FPScreen is efficient, and the results are more accurate.

Subsequently, we plan to continue to improve our search tools and methods from three aspects. 1) Continue to improve the search efficiency; 2) optimize the input mode of the target molecule; 3) improve the search accuracy, such as improving the number of features of molecular fingerprints.